\documentclass[%
 aip,
 amsmath,amssymb,
reprint,%
]{revtex4-2}

\usepackage{graphicx}
\usepackage{dcolumn}
\usepackage{bm}

\usepackage[utf8]{inputenc}
\usepackage[T1]{fontenc}
\usepackage{mathptmx}

\begin{document}


\title{Ultra-short pulse generation from mid-IR to THz range using plasma wakes and relativistic ionization fronts}

\author{Zan Nie}
 \email{znie@ucla.edu}
\author{Yipeng Wu}
\author{Chaojie Zhang}
\affiliation{ 
Electrical and Computer Engineering Department, University of California Los Angeles, Los Angeles, California 90095, USA
}

\author{Warren B. Mori}
\affiliation{ 
Electrical and Computer Engineering Department, University of California Los Angeles, Los Angeles, California 90095, USA
}
\affiliation{%
Department of Physics, University of California Los Angeles, Los Angeles, California 90095, USA
}%

\author{Chan Joshi}
\affiliation{ 
Electrical and Computer Engineering Department, University of California Los Angeles, Los Angeles, California 90095, USA
}

\author{Wei Lu}
\author{Chih-Hao Pai}
\author{Jianfei Hua}
\affiliation{ 
Department of Engineering Physics, Tsinghua University, Beijing, 100084, China
}

\author{Jyhpyng Wang}
\affiliation{ 
Department of Physics, National Central University, Jhongli 32001, Taiwan
}
\affiliation{ 
Institute of Atomic and Molecular Sciences, Academia Sinica, Taipei 10617, Taiwan
}
\affiliation{ 
Department of Physics, National Taiwan University, Taipei 10617, Taiwan
}

\date{\today}

\begin{abstract}
This paper discusses numerical and experimental results on frequency downshifting and upshifting of a 10\,$\mu$m infrared laser to cover the entire wavelength (frequency) range from $\lambda$=1-150\,$\mu$m ($\nu$=300-2\,THz) using two different plasma techniques. The first plasma technique utilizes frequency downshifting of the drive laser pulse in a nonlinear plasma wake. Based on this technique, we have proposed and demonstrated that in a tailored plasma structure multi-millijoule energy, single-cycle, long-wavelength IR (3-20\,$\mu$m) pulses can be generated by using an 810\,nm Ti:sapphire drive laser. Here we extend this idea to the THz frequency regime. We show that sub-joule, terawatts, single-cycle terahertz (2-12 THz, or 150-25\,$\mu$m) pulses can be generated by replacing the drive laser with a picosecond 10\,$\mu$m CO$_2$ laser and a different shaped plasma structure. The second plasma technique employs frequency upshifting by colliding a CO$_2$ laser with a rather sharp relativistic ionization front created by ionization of a gas in less than half cycle (17\,fs) of the CO$_2$ laser. Even though the electrons in the ionization front carry no energy, the frequency of the CO$_2$ laser can be upshifted due to the relativistic Doppler effect as the CO$_2$ laser pulse enters the front.  The wavelength can be tuned from 1-10\,$\mu$m by simply changing the electron density of the front. While the upshifted light with $5 <\lambda(\mu$m$)< 10$ propagates in the forward direction, that with $1 <\lambda(\mu$m$)< 5$ is back-reflected. These two plasma techniques seem extremely promising for covering the entire molecular fingerprint region.    
\end{abstract}

\maketitle

\section{Introduction}

Ultra-short radiation source in the mid-infrared (mid-IR) to terahertz (THz) range (1–150\,$\mu$m or 300–2\,THz) is highly desirable in numerous physics, material science, and biology applications \cite{Meckel2008, Forst2011, Popmintchev2012, Ghimire2011, Wolter2015, Vampa2015, Schubert2014, Hassan2016, Pupeza2020}. These ultra-short mid-IR/THz sources are ideal tools for pump-probe experiments in the “molecular fingerprint” region \cite{Meckel2008, Forst2011}, ultrafast X-ray high harmonic generation (HHG)\cite{Popmintchev2012}, coherent IR spectroscopy \cite{Auton1985, Matsubara2013}, time-resolved imaging of molecular structures \cite{Blaga2012}, and nanotip photoemission \cite{Wimmer2014}. Furthermore, intense single-cycle mid-IR/THz pulses are very useful in attosecond science applications, for instance, sub-femtosecond control and metrology of bound-electron dynamics in atoms \cite{Hassan2016}, generation of attosecond \cite{Silva2015} or even zeptosecond pulses \cite{Hernandez-Garcia2013}, and coherent control of lattice displacements through nonlinear phononics \cite{Forst2011}. Remarkable progress for generating such mid-IR/THz pulses has been made through various optical methods in nonlinear crystals or gases, such as optical parametric amplification (OPA) \cite{Fu2018,Sanchez2016}, difference-frequency generation (DFG) \cite{Pupeza2015,Krogen2017, Novak2018, Vicario2014}, optical rectification \cite{Sell2008,Junginger2010}, and two-color filamentation \cite{Fuji2015,Yoo2017}. Recent numerical and experimental studies show that plasmas can also be utilized as nonlinear optical media to generate ultra-short mid-IR/THz pulses with high efficiency \cite{Nie2018,Nie2020}. Unlike in traditional nonlinear crystals, there is no damage concerns for the power/intensity scaling in plasmas, which makes it possible to generate relativistically intense pulses in the mid-IR to THz spectral range.

In this paper we discuss numerical and experimental results on two different plasma techniques for frequency downshifting and upshifting of an IR laser to cover the entire mid-IR to THz band that covers wavelengths from 1-150\,$\mu$m (or frequencies from 300-2 THz). In Sec.\,\ref{sec2} we briefly describe the first plasma technique that uses frequency downshifting (or photon deceleration) in a nonlinear plasma wake. We first summarize our particle-in-cell (PIC) simulations and experiments on generating multi-millijoule, single-cycle pulses in the 3-20\,$\mu$m wavelength range with a specially prepared plasma density structure \cite{Nie2018,Nie2020}. In Sec.\,\ref{sec3} we extend this frequency downshifting scheme from LWIR to THz spectral range by replacing the drive laser with a picosecond 10\,$\mu$m CO$_2$ laser. In this case we have to use a different profile of the plasma density structure. By PIC simulations we show that sub-joule, terawatts, single-cycle THz pulses can be produced using a picosecond 10\,$\mu$m CO$_2$ drive laser. In Sec.\,\ref{sec4} we introduce the second plasma technique that frequency upshifts the incident IR laser pulse using a relativistic ionization front and show corresponding numerical results. In this scheme, a long-wavelength (e.g. 10\,$\mu$m CO$_2$ laser) pulse collides with a relativistic ionization front produced by photoionization of a column of gas using a short-wavelength (e.g. 800\,nm Ti:sapphire laser) pulse. The frequency change of the CO$_2$ laser pulse and the direction in which the upshifted radiation is observed is obtained by the double relativistic Doppler shift formulism using Lorentz transformations. It is found that the frequency can be tuned by simply tuning the gas and hence the plasma density.

\section{\label{sec2}Extreme Frequency downshifting in a nonlinear plasma wake (mid-IR to LWIR range)}
\begin{figure}
\includegraphics[width=0.45\textwidth]{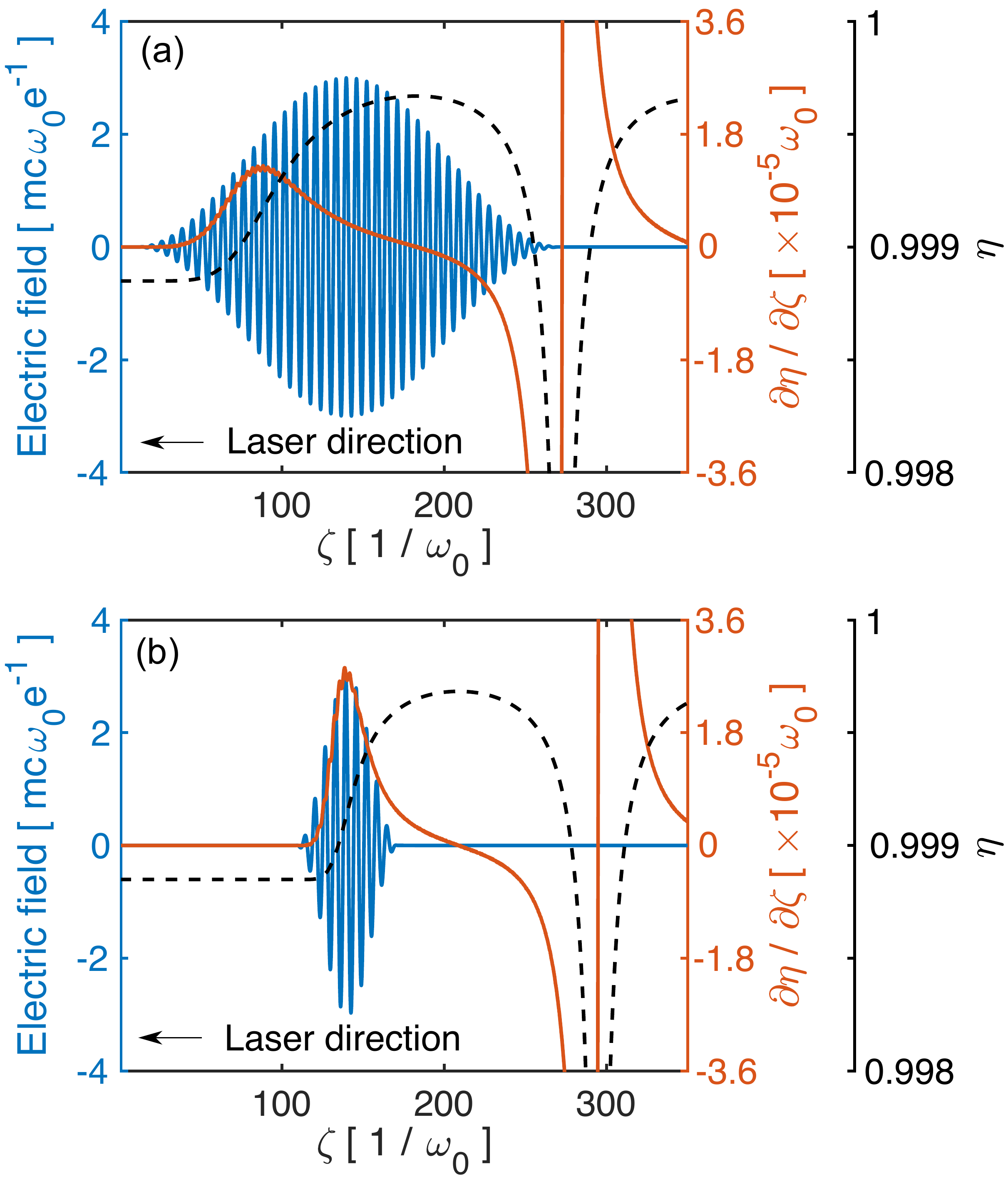}
\caption{\label{fig1} 
Comparison of frequency downshifting for two different pulse durations. The plasma density is $4\times 10^{18}\,\text{cm}^{-3}$. The normalized vector potential is 3. The pulse duration is 40.0\,fs in (a) and 9.6\,fs in (b). The blue line, orange line, and dashed black line show the laser electric field, the refractive index gradient, and the refractive index, respectively.
}
\end{figure}

It is known that an ultra-short intense laser pulse can excite a nonlinear wake (density disturbance) while propagating through an underdense ($\omega_p<\omega_0$) plasma. Here $\omega_p=(4\pi n_p e^2/m)^\frac{1}{2}$ is the plasma frequency, $n_p$ is the plasma density, $m$ is the electron mass, $e$ is the electron charge, and $\omega_0$ is the laser frequency. If the laser’s normalized vector potential $a_0=eE/m\omega_0 c>2$ ($E$ is the electric field of the laser pulse and $c$ is the speed of light) and the pulse duration $\tau<\sqrt{2\pi /\omega_p}$, the ponderomotive force of the laser pulse eventually pushes out all the plasma electrons forward and outward, forming a 3D nonlinear wake \cite{Lu2006}. According to 1D nonlinear theory, the refractive index seen by the co-propagating laser photons varies as \cite{Sprangle1990,Sprangle1990a}: $\eta \simeq 1-\frac{\omega_p^2}{2\omega_0^2} \frac{1}{1+\phi}$, where $\phi=|e|\Phi/m c^2$ is the normalized scalar potential. The refractive index gradient $\frac{\partial\eta}{\partial \zeta}$ with $\zeta=t-z/c$ being the variable in the speed of light frame continuously alters the photon frequency via self-phase modulation (SPM) such that the instantaneous frequency is given by $\omega(t)=\omega_0-\omega_0 \int{\frac{\partial\eta}{\partial \zeta} dt}$ (or $\frac{1}{\omega} \frac{\partial \omega}{\partial t} =-\frac{\partial\eta}{\partial \zeta}$). Specifically, photons in the front of the wake where  $\frac{\partial\eta}{\partial \zeta}$ is positive are frequency-downshifted, while photons in the tail of the wake where  $\frac{\partial\eta}{\partial \zeta}$ is negative are frequency-upshifted, and the part in the central, near electron-free region experiences scarcely any frequency change (Fig.\,\ref{fig1}). Due to the negative group velocity dispersion (GVD) of the plasma, the longer wavelength photons generated by a positive refractive index gradient travel with a smaller group velocity ($v_g(\omega)\simeq c[1-\frac{\omega_p^2}{2\omega^2}]$) than the shorter wavelength photons. This is the reason why this phenomenon is also called photon acceleration/deceleration \cite{Sprangle1990, Sprangle1990a, Wilks1989,Esarey1990,Mori1997}. This mechanism has been used for pulse compression \cite{Tsung2002, Gordon2003,Faure2005} or as a diagnostic for the generation of wakes in plasmas \cite{Downer2018,Murphy2006,Shiraishi2013}. 

When the laser pulse duration is roughly one plasma wavelength ($c\tau\simeq \lambda_\text{p}$, Fig.\,\ref{fig1}(a)), laser photons are frequency downshifted at the front and upshifted at the back of the pulse just as in usual SPM in fibers or neutral gases. However, when the laser pulse duration is much less than one plasma wavelength ($c\tau\ll \lambda_\text{p}$, Fig.\,\ref{fig1}(b)), the whole pulse resides at the very front of the wake and experiences only frequency downshifting (photon deceleration). This is so-called asymmetric SPM. Using this asymmetric SPM concept, the possibility of frequency downshifting near-IR drive pulses to mid-IR range in uniform-density plasmas was explored numerically \cite{Zhu2012,Zhu2013} and experimentally \cite{Pai2010,Schreiber2010,Streeter2018} in the last decade.

Recently, we have extended this concept to the LWIR range by using a tailored plasma structure \cite{Nie2018,Nie2020}. Using plasma theory, we found that there is an optimal pulse duration for efficient frequency downshifting\cite{Nie2018} expressed as $c\tau \simeq \frac{0.52\lambda_\text{p}}{a_0}$, where $\lambda_\text{p}$ is the plasma wavelength. In such an optimal case, the longitudinal profile of the laser pulse roughly overlaps with the longitudinal profile of the refractive index gradient so that the whole pulse experiences frequency downshifting with the optimum efficiency (Fig.\,\ref{fig1}(b)). For example, if $a_0=3$, $n_\text{p}=4\times 10^{18}\,\text{cm}^{-3}$, then the optimal pulse duration is 9.6 fs, which is the case shown in Fig.\,\ref{fig1}(b). Such short ($\sim$10\,fs) pulses with high intensity ($a_0\sim 3$) are very difficult to obtain even with the state-of-the-art techniques. To tackle this issue, we put forward a general solution that uses a tailored plasma structure that first compresses and then frequency downshifts commonly available longer ($\sim$30\,fs) Ti:sapphire laser pulses to the LWIR range \cite{Nie2018}. The plasma structure consists of two sections: a relatively low-density but longer uniform density region connected to a higher density but shorter density region by a sharp density up-ramp. The 3D OSIRIS \cite{Fonseca2002,Fonseca2008} simulation showed that the originally 30\,fs drive pulse is first compressed to $\sim$10\,fs (the optimal pulse duration for following frequency downshifting) in the first section, and then experiences rapid and efficient frequency downshifting in the second section. Using such a tailored plasma structure, multi-millijoule, near-single-cycle, relativistic pulses in the LWIR region are generated~\cite{Nie2018}. The central wavelength tunability in the range of 5-14\,$\mu$m is achieved by varying the peak density in the short high-density region. 

\begin{figure}
\includegraphics[width=0.4\textwidth]{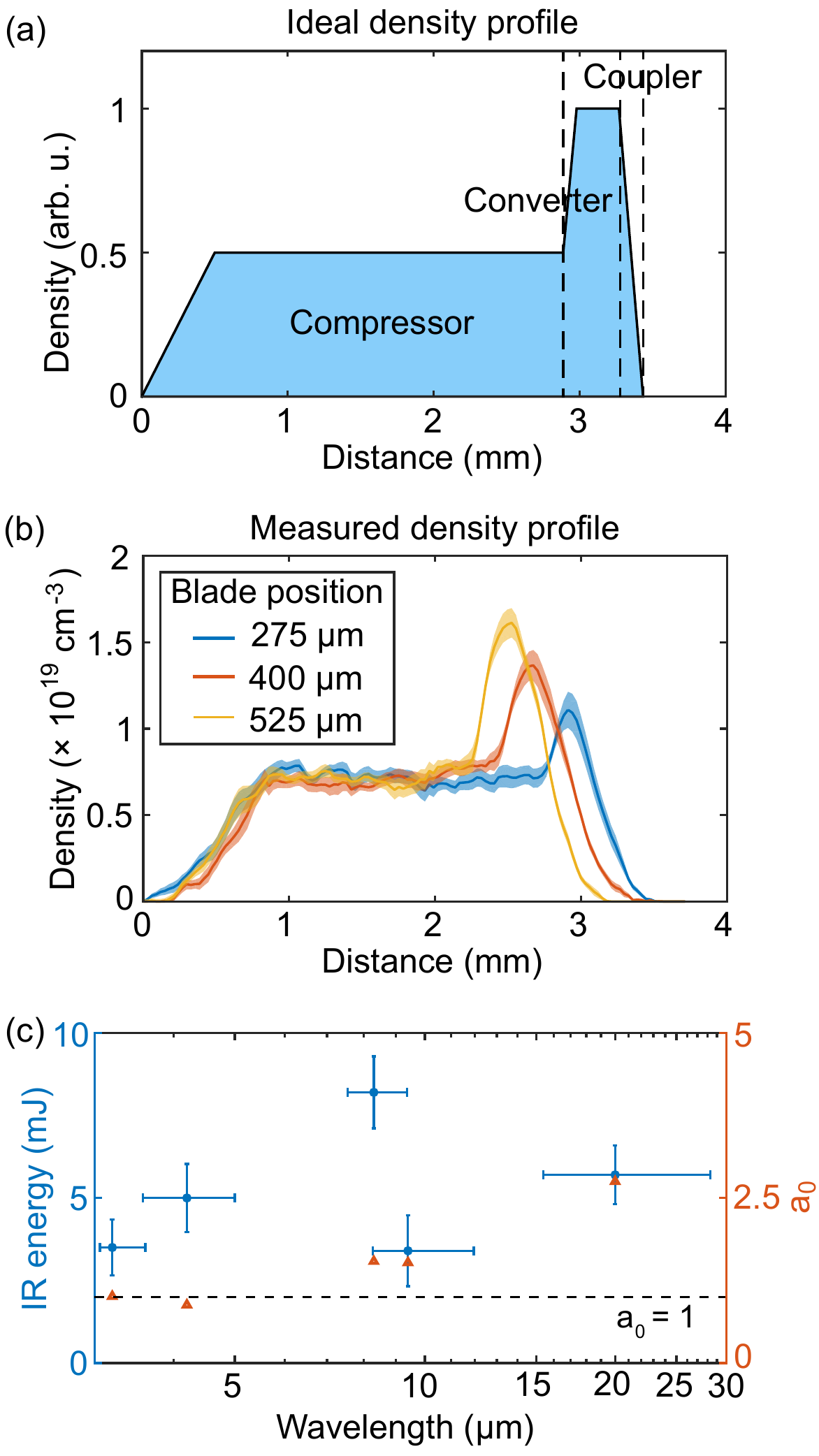}
\caption{\label{fig2} 
(a) Ideal plasma density profile for efficient frequency downshifting. (b) Measured plasma density profile produced by a movable blade covering a fraction of a gas jet. (c) Wavelength tunability in the range of 3-20\,$\mu$m. Reproduced with permission from Nie et al., Nat. Photonics 12, 489-494 (2018) (Copyright 2018 Springer Nature) and Nie et al., Nat. Comms. 11, 2787 (2020) (Copyright 2020 The Authors).
}
\end{figure}

We have experimentally demonstrated this frequency downshifting scheme by employing a tailored plasma structure scheme\cite{Nie2020}. A supersonic hydrogen gas jet target with an insertable blade was used to shock-induce a density spike \cite{Schmid2010,Gonsalves2011,Buck2013} in the gas flow to produce a similar density profile as we have described above as shown in Fig.\,\ref{fig2}(a) and (b). In the experiment, an ultra-short (36$\pm$2\,fs FWHM), energetic (580$\pm$9\,mJ, 16\,TW), 810\,nm wavelength drive laser pulse \cite{Hung2014} passed through this tailored plasma structure, fully ionized the neutral gas and produced a highly nonlinear wake \cite{Lu2006}. The strong time-dependent plasma density gradients formed during the expulsion of the plasma electrons phase modulated the laser pulse downshifting the instantaneous frequency of the photons as explained earlier. The generated LWIR pulse was characterized by cross-correlation frequency-resolved optical gating (XFROG) \cite{Linden1998} method based on four-wave mixing (FWM) in argon gas. By optimizing the peak density of the sharp high-density section of the plasma structure, we have generated 3.4$\pm$1.1\,mJ, 32.0\,fs (FWHM), LWIR pulses with a central wavelength of 9.4\,$\mu$m, demonstrating the generation of relativistic, near single-cycle LWIR pulses. We also showed that the central wavelengths could be tuned in the range of 3–20\,$\mu$m by varying several experimental parameters, such as the gas density profile and laser energy. These data are summarized in Fig.\,\ref{fig2}(c) where we show the measured energy and the deduced peak $a_0$ of the frequency downshifted light from 3-20\,$\mu$m range.

\section{\label{sec3}Frequency downshifting in a nonlinear plasma wake (THz range)}
Now we extend this idea to obtain energetic THz radiation. Although research on generating intense ultra-short THz pulses has made great strides in the past decade, one of the current challenges facing THz technology is how to generate sub-joule, terawatts level, single-cycle THz pulses. Most existing methods, such as DFG \cite{Vicario2014}, optical rectification \cite{Sell2008,Junginger2010} and two-color filamentation \cite{Yoo2017}, rely on frequency conversion from commonly used near-infrared lasers. The major issue of these methods is the low conversion efficiency. One straightforward way to improve this problem is to use longer-wavelength drive laser. CO$_2$ laser is currently the most powerful LWIR ($\sim$10\,$\mu$m) laser close to THz spectral range. In recent years, multi-terawatt picosecond 10\,$\mu$m CO$_2$ lasers have been developed \cite{Haberberger2010,Polyanskiy2011}, that  provides a better alternate path to nonlinear optics for generating intense ultra-short THz pulses. 

The concept of frequency downshifting in plasmas described in Sec.\,\ref{sec2} can be extended to THz range if switching the drive lasers from Ti:sapphire lasers ($\sim$800\,nm) to CO$_2$ lasers ($\sim$10\,$\mu$m). Based on this idea, sub-joule, terawatts, single-cycle THz pulses can be generated by using a picosecond 10\,$\mu$m CO$_2$ laser to excite a nonlinear wake in a tailored plasma structure described in Sec.\,\ref{sec2}; however, the details of the plasma density structure are different. 

\begin{figure*}
\includegraphics[width=0.8\textwidth]{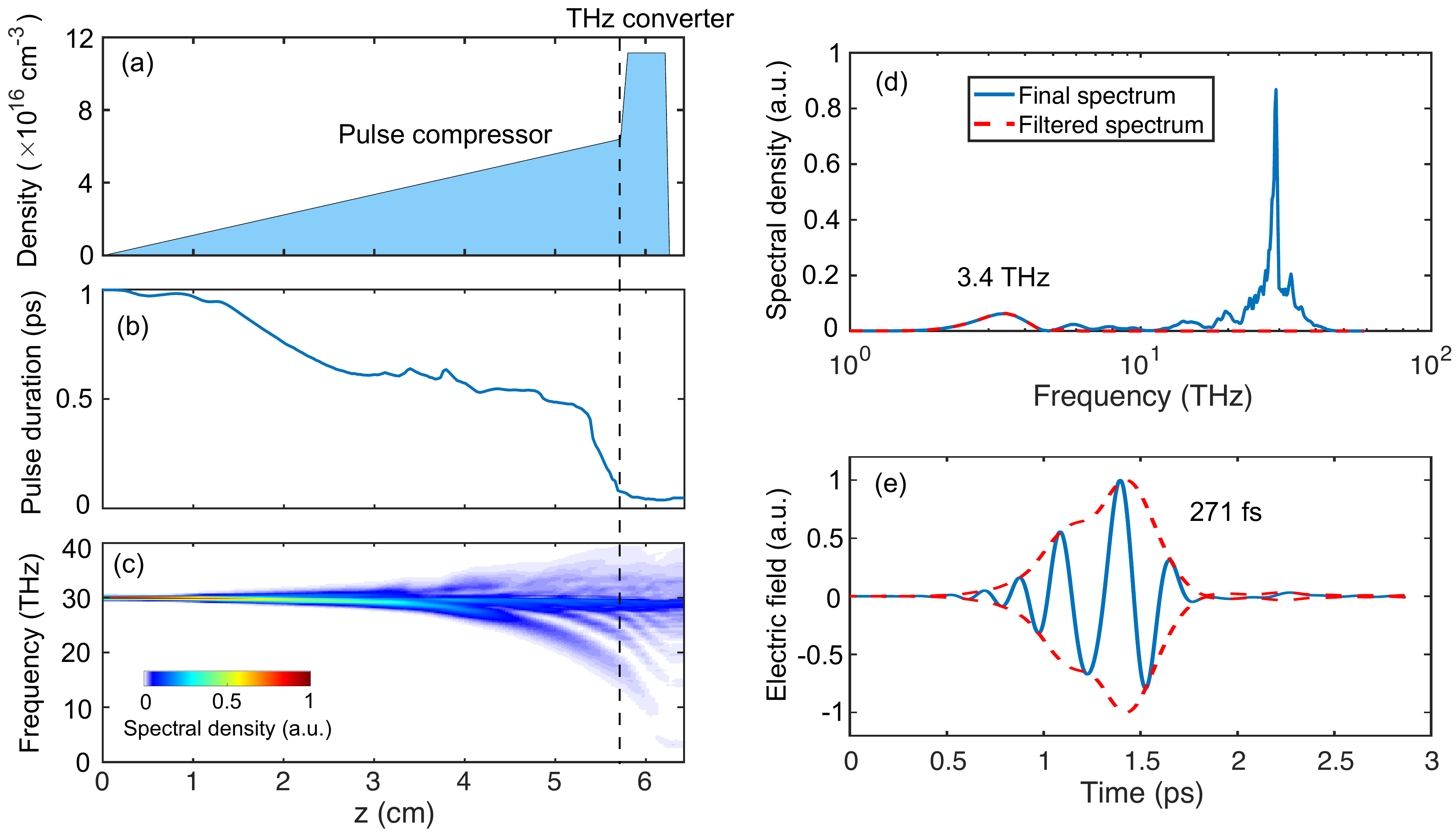}
\caption{\label{fig3} 
Quasi-3D OSIRIS simulation results on generation of sub-joule, terawatts, single-cycle THz pulses. (a) The tailored plasma density profile. (b,c) Evolution of pulse duration and spectrum with propagation distance in the plasma. (d) Final spectrum and filtered single-cycle THz spectrum. (e) The filtered single-cycle THz pulse.
}
\end{figure*}

\begin{figure*}
\includegraphics[width=0.9\textwidth]{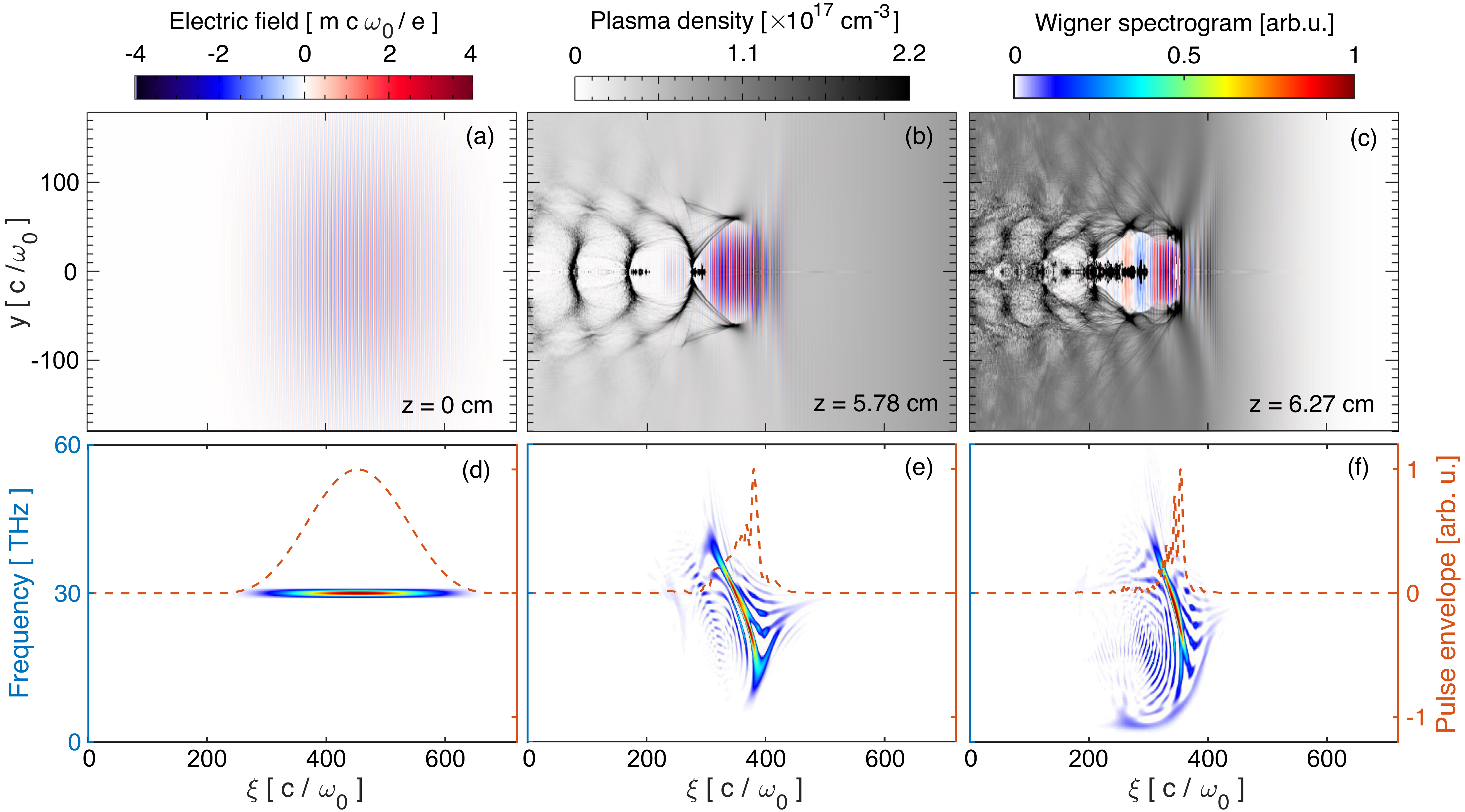}
\caption{\label{fig4} 
Three snapshots showing the evolution of the laser pulse and wakefield. (a-c) The transverse electric field of the laser pulse and the plasma density distribution (a) at the beginning of the plasma structure , (b) at the end of pulse compressor,  and (c) at the end of THz converter. (d-f) The Wigner spectrograms of the on-axis transverse electric field and laser envelope (orange dashed line) at the same position as in (a-c), respectively.
}
\end{figure*}

Here, we present the PIC simulation results calculated by Quasi-3D OSIRIS code \cite{Lifschitz2009,Davidson2015}. The drive CO$_2$ laser pulse has an energy of 14.6\,J with pulse duration of 1\,ps (FWHM). It is focused to a spot size of $w_0=130.5\,\mu$m at the front of the tailored plasma structure.  At the initial $a_0$ of 2, the laser spot size is matched to the plasma at a density of $1.3\times 10^{16}\,\text{cm}^{-3}$. However, with the increase of the plasma density, wake excitation leads to self-compression and self-focusing which increases the $a_0$ and decreases the laser spot size. In our simulation, the laser beam is seen to be self-guided in the wake throughout the whole propagation distance \cite{Ralph2009}. As discussed in the previous case, the tailored plasma structure in this case (Fig.\,\ref{fig3}(a)) also consists of two sections: the pulse compressor and the THz converter. The pulse compressor section is a long plasma up-ramp, which is different than the uniform density (flattop) profile of the pulse compressor section in the tailored structure (Fig.\,\ref{fig2}(a) and (b)) proposed in Sec.\,\ref{sec2}. The reason is that the normalized vector potential $a_0$ is lower and the pulse length (scaled to drive laser wavelength) is longer in this CO$_2$ drive laser case compared with the previous Ti:sapphire drive laser cases, which makes it more challenging to compress the CO$_2$ pulse. The long up-ramp plasma profile is designed to compress such a 1-ps drive pulse in an as short a distance as possible. At the beginning of the plasma, the plasma density should be low enough so that the plasma wavelength is longer than the original pulse length to avoid pulse splitting by overlapping with two wake buckets. As the laser pulse self-compresses, the plasma density increases accordingly to speed up pulse shortening and minimize the compression distance. In this way, such an up-ramp plasma profile compresses the originally 1-ps pulse to less than 100\,fs (Fig.\,\ref{fig3}(b), Fig.\,\ref{fig4}(b,e)).

Then in the THz converter section, the compressed drive laser is rapidly frequency-downshifted in a higher-density plasma region to generate the THz photons. The THz photons then slip backwards towards the center of the wake due to their much smaller group velocity. The central region of the wake is nearly devoid of plasma electrons ($\eta\sim1$) and thus serves as a perfect container for the THz pulse (Fig.\,\ref{fig4}(c,f)). Finally, the generated THz pulse has a pulse duration of 271\,fs (FWHM), corresponding to a single-cycle pulse for the central frequency of 3.4 THz. The estimated THz energy (1-5\,THz) is 293\,mJ, with conversion efficiency of $\sim$2\,\% and peak power as high as 1.1\,TW. Accordingly, the peak field of the THz pulse at the exit of plasma structure reaches as high as 2.3\,GV/cm.

\begin{figure}
\includegraphics[width=0.35\textwidth]{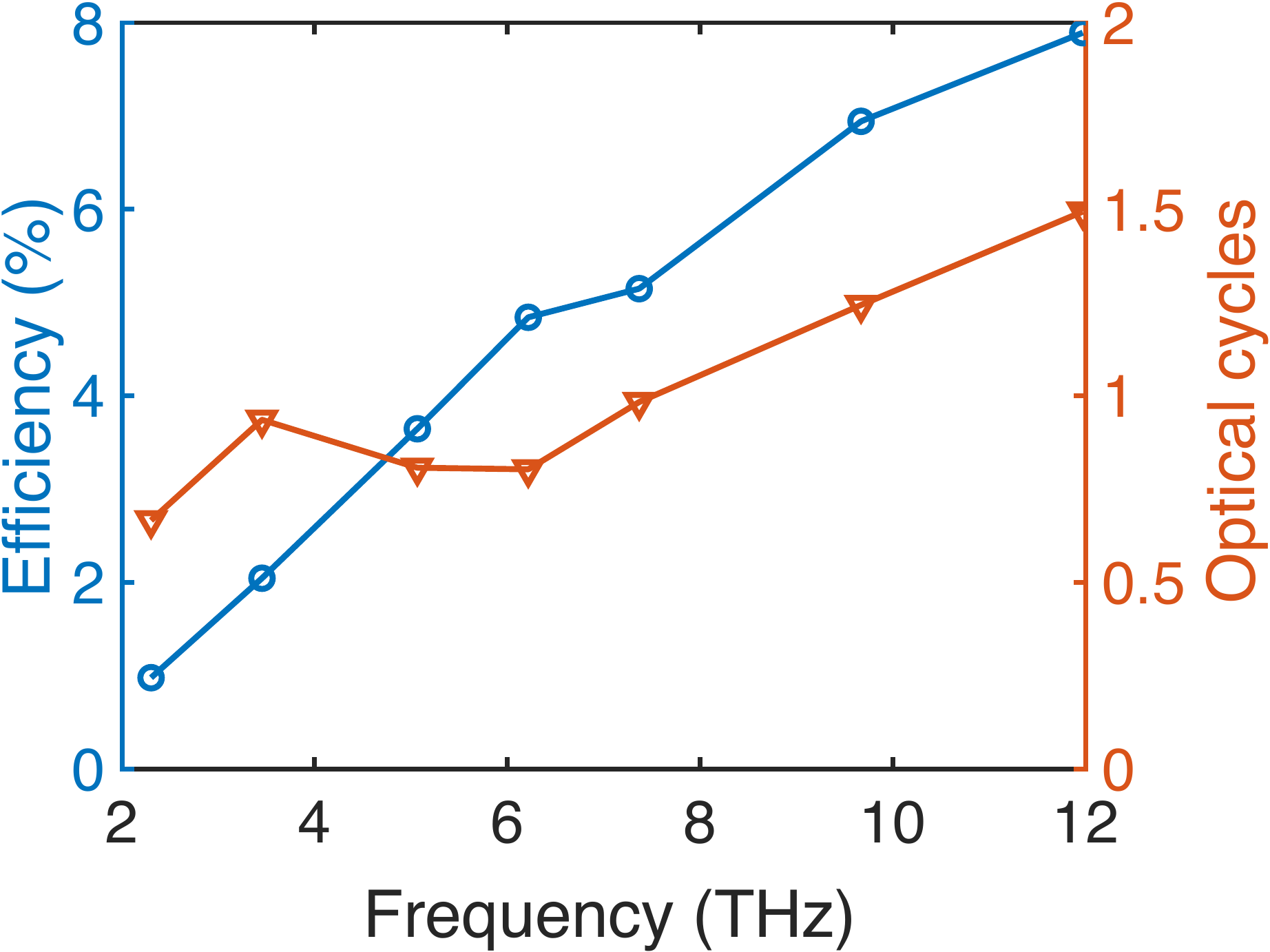}
\caption{\label{fig5} 
Conversion efficiency of frequency downshifting and optical cycles v.s. output central frequency.
}
\end{figure}

Furthermore, the output THz frequency can be tuned from 2-12\,THz by varying the plasma density and length in the THz converter section while keeping other parameters unchanged. From Fig.\,\ref{fig5}, one can see that the conversion efficiency roughly linearly scales with the output frequency, and the THz pulses in different frequency cases remain pulse duration at around single cycle or even sub cycle at lower frequency cases.

Although the driving laser pulse duration used here (1\,ps) is at present less than the shortest available CO$_2$ pulse duration (2\,ps)\cite{Polyanskiy2020}, a great deal of work is being carried out to shorten CO$_2$ pulses to 1-ps or even sub-ps pulse durations in the near future \cite{Panagiotopoulos2020,Tovey2019}. Even if using the currently available 2-ps CO$_2$ laser pulse, we can still use the same method with some adaptation. A plasma channel may be used to confine the laser beam in long propagation distance of the pulse compressor section, since the plasma density in the pulse compressor has to be lower and self-guiding at lower densities may not be possible.

\section{\label{sec4}Frequency upshifting by a relativistic ionization front}
Now, we introduce another plasma technique: frequency upshifting by a relativistic ionization front. This mechanism was first studied theoretically \cite{Lampe1978,Mori1991}, and then observed using microwave radiation colliding with an overdense ionization front \cite{Savage1992,Lai1996}. Here, we propose using this scheme to frequency upshift a 10\,$\mu$m CO$_2$ laser by colliding it with a relativistic ionization front produced by a short-wavelength ionizing laser pulse.

\begin{figure}
\includegraphics[width=0.45\textwidth]{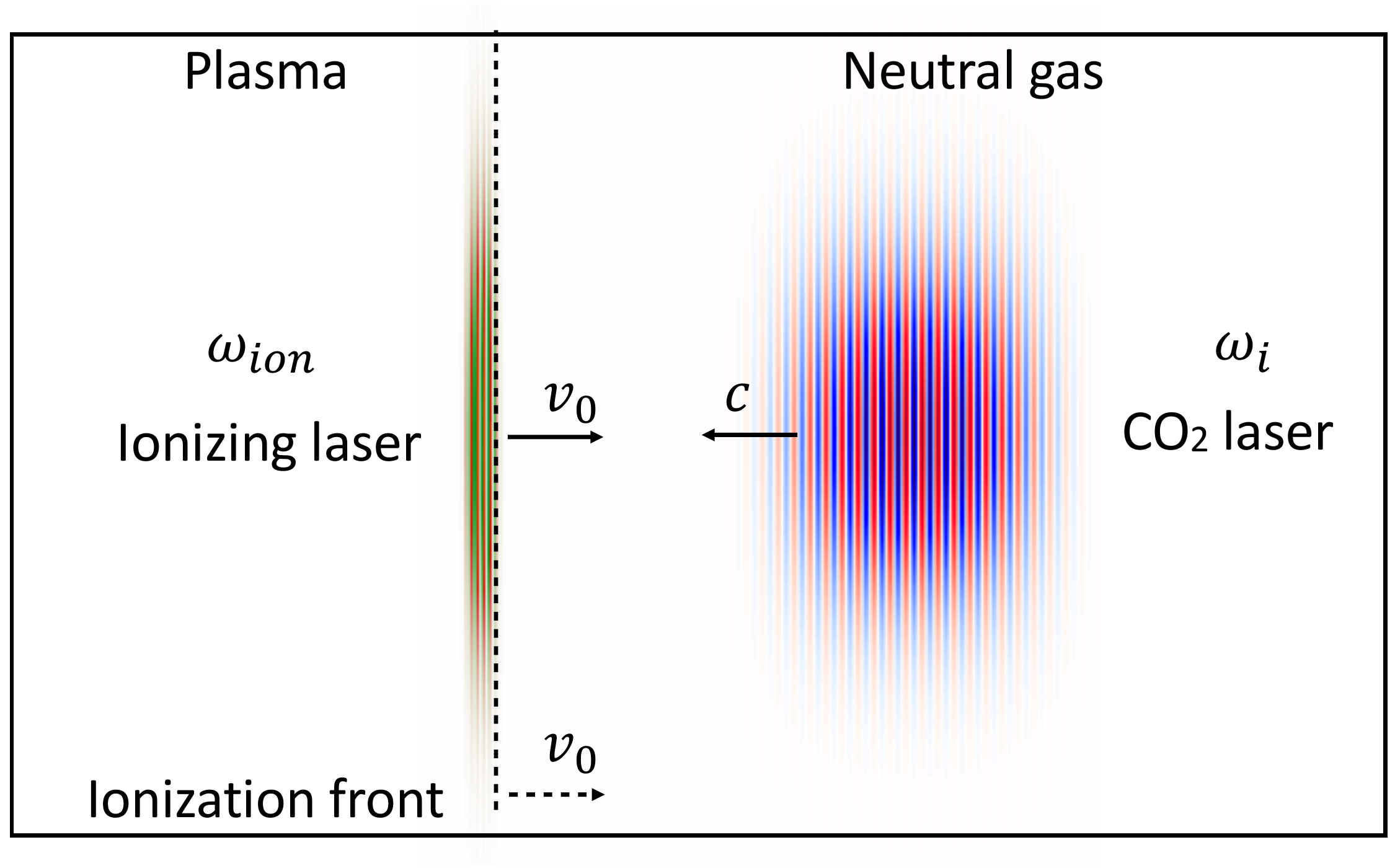}
\caption{\label{fig6} 
Schematic of frequency upshifting of a CO$_2$ laser pulse by a relativistic ionization front produced by a short-wavelength ionizing laser.
}
\end{figure}

The proposed schematic is shown in Fig.\,\ref{fig6}. The ionizing laser with a frequency of $\omega_{ion}$ propagates into the neutral gas. It ionizes the neutral gas and produces a relativistic ionization front. To the lowest order, the front’s velocity $v_0$ is the group velocity of the ionizing laser in plasma. The CO$_2$ laser with a frequency of $\omega_i$ ($\omega_i \ll \omega_{ion}$) counter-propagates with the ionizing laser and collides with the ionization front. Then the frequency of the CO$_2$ laser is upshifted due to relativistic Doppler effect. Here, we use Lorentz transformations to investigate this problem. Switching from the laboratory frame to the ionization front frame, the incident CO$_2$ laser frequency of $\omega_i$ will be upshifted to $\omega_f\simeq 2\gamma_0 \omega_i$, where $\gamma_0=(1-\frac{v_0^2}{c^2} )^{-\frac{1}{2}}$. Depending on the plasma density of the ionization front, there are two possible regimes \cite{Mori1991}. When the plasma density is high enough so that $\omega_p > \omega_f$ ($\omega_p^2 > 2\omega_{ion}\omega_i$), it is in the overdense regime. The CO$_2$ laser will be reflected by the ionization front, and the frequency of the reflected wave in the lab frame is $\omega_r\simeq 4\gamma_0^2 \omega_i$.

When the plasma density $\omega_p < \omega_f$ ($\omega_p^2 < 2\omega_{ion}\omega_i$), it is in the underdense regime. The CO$_2$ laser will enter the ionization front, and the frequency of the transmitted wave is now $\omega_t=\gamma_0 (\omega_f-v_0 k_{tf})$, where $k_{tf}=\frac{1}{c} \sqrt{\omega_f^2-\omega_p^2}$ is the wavenumber of the transmitted wave in the ionization front frame. The transmitted frequency increases with the increase of the plasma density and is bounded by $\omega_{ion}$ (when $\omega_p^2 = 2\omega_{ion}\omega_i$). In the underdense limit ($\omega_p^2 \ll 2\omega_{ion}\omega_i$), the transmitted frequency and wavenumber are approximately $\omega_t\simeq \omega_i (1+\frac{\omega_p^2}{4\omega_i^2})$ and $k_t\simeq \frac{\omega_i}{c} (1-\frac{\omega_p^2}{4\omega_i^2})$. Interestingly, the transmitted wavenumber could be positive, zero, or negative. When the wavenumber is zero, the group velocity of the wave goes to zero and the wave will be absorbed by the plasma. When the wavenumber is negative, the wave will travel backwards in the laboratory frame, moving with a group velocity less than the velocity of the ionization front so that it is still transmitted in the ionization front frame. 

As observed in a previous study \cite{Mori1991}, the underdense regime instead of the overdense regime is recommended in practical applications for two reasons. One is that the frequency upshift in the underdense regime is independent of $\gamma_0$, therefore not affected by the potential variation of $\gamma_0$ due to the variation of the group velocity of ionizing laser when ionizing the gas. The other advantage is that the plasma density needed is lower than the overdense case for the same incident and upshifted frequencies.

\begin{figure}
\includegraphics[width=0.5\textwidth]{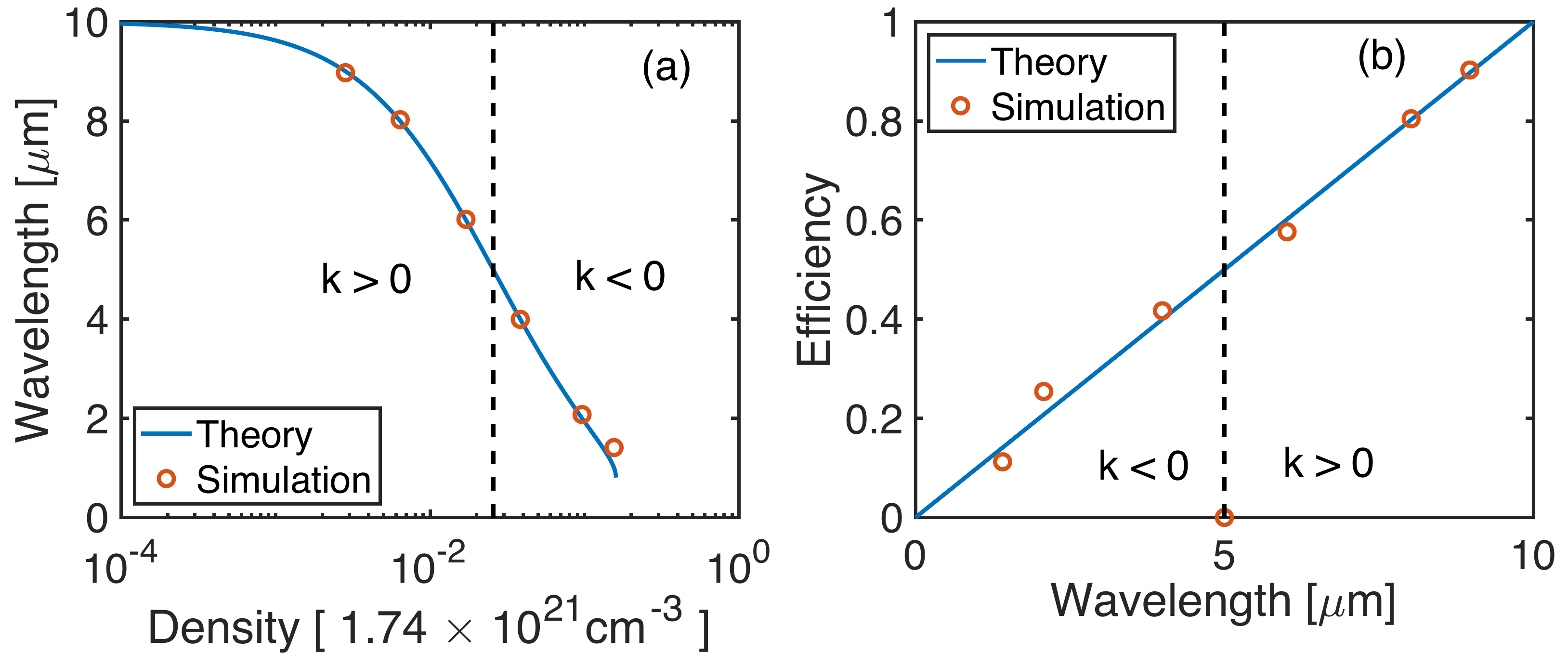}
\caption{\label{fig7} 
Frequency upshifting in the underdense regime. (a) Output wavelength v.s. ionization front density. (b) Energy conversion efficiency v.s. output wavelength. The dashed line marks the boundary of $k = 0$.
}
\end{figure}

We have run a series of 2D OSIRIS\cite{Fonseca2002,Fonseca2008} simulations to show frequency tuning dependence on plasma density in the underdense regime. In our simulations, the ionizing laser is a 40-fs 800-nm Ti:sapphire laser with a focused spot size of 200\,$\mu$m and normalized vector potential $a_0=0.05$. The CO$_2$ laser pulse duration is 333\,fs here (to save simulation time) with a focused spot size of 100\,$\mu$m and normalized vector potential $a_0=0.05$. Hydrogen is used as the neutral gas. The Ti:sapphire laser can ionize the hydrogen gas but the CO$_2$ laser cannot since its peak intensity is below the tunneling ionization threshold intensity for hydrogen. The output wavelength can be tuned in the spectral range of 1-10\,$\mu$m by simply tuning the gas density as shown in Fig.\,\ref{fig7}(a). In the underdense limit the reflected wave is very weak, thus transmission coefficient in the laboratory frame approaches unity and the pulse duration will be compressed due to the conservation of oscillation cycles \cite{Mori1991} if not considering later-on pulse stretching by dispersion. Therefore, the energy conversion efficiency approximately equals to the ratio of the output wavelength and the initial wavelength ($\eta\simeq\lambda_t/\lambda_i$) as shown in Fig.\,\ref{fig7}(b). 

For the sake of completeness, we should also mention that physically it is the boundary conditions on $\bold{E}$ and $\bold{H}$ at the ionization front that give rise to the reflected and transmitted waves. In addition, the plasma ionization front can support currents. If the onset of ionization is sudden, the ionization front width is much less than the wavelength of the CO$_2$ . The plasma electrons are produced at rest, therefore the current density $\bold{J}=0$. Thereafter as the ionization front passes through the linearly polarized CO$_2$ laser pulse, its electric field gives rise to sinusoidal oscillations of current that in turn give rise to regions of static magnetic field at half the wavelength of the CO$_2$ laser. This so-called magneto-static mode \cite{Mori1991,Fiuza2010} is the third natural mode in an unmagnetized plasma in addition to the Bohm-Gross waves and ion acoustic waves. This mode has not been conclusively identified to date in plasmas because of the difficulty of producing a periodic arrangement of quasi-dc currents in plasmas. The interaction of an ionization front with a counter propagating electromagnetic wave as described here will allow one to generate the conditions necessary to excite this mode in addition to frequency upshifted light that is the topic of this paper.

\section{Conclusions}
In summary, we show in this paper that by using two different plasma techniques we can generate mid-IR/THz pulses that cover the entire bandwidth from 1-150\,$\mu$m (300-2 THz).

By using frequency downshifting of a Ti:sapphire laser in a nonlinear wake, the generation of relativistic, near single-cycle LWIR pulses tunable in 3-20\,$\mu$m range has already been demonstrated using a tailored plasma structure. Given that a few TW class, femtoseconds drive lasers in the near-IR are now commonplace, one expects this technique to be adopted in many laboratories to give intense, tunable LWIR pulses. Further extending this scheme into THz range is achieved by using a picosecond CO$_2$ driving laser instead of a Ti:sapphire laser. By PIC simulations, we show that sub-joule, terawatts, single-cycle THz pulses are generated and can be frequency tuned in the range of 2-12 THz. At relativistic intensities afforded by such mid-IR/THz sources it will now be possible to study laser wakefield and direct laser acceleration by long-wavelength drivers not only in plasmas but also in dielectric and semiconductor structures.

By using frequency upshifting of a CO$_2$ laser by colliding it with a relativistic ionization front, we show by PIC simulations that mid-IR pulses tunable in the spectral range of 1-10\,$\mu$m can be produced. The tunability is straightforward by simply tuning the gas density. Such mid-IR pulses with relatively lower intensities can be used for pump-probe experiments in the molecular fingerprint region\cite{Meckel2008, Forst2011}, high harmonic generation\cite{Popmintchev2012, Ghimire2011, Vampa2015}, and resonant or non-resonant nonlinear interactions in gases, solids, or biological systems\cite{Schubert2014, Hassan2016, Pupeza2020}.

\begin{acknowledgments}
This work was supported by the Office of Naval Research (ONR) MURI (N00014-17-1-2075), AFOSR grant FA9550-16-1-0139, U.S. Department of Energy grant DE-SC0010064 and NSF grant 1734315. The simulations were performed on Sunway TaihuLight, Hoffman cluster at UCLA, and NERSC at LBNL.
\end{acknowledgments}

\section*{Data availability}
The data that support the findings of this study are available from the corresponding author upon reasonable request. 

\section*{References}
\nocite{*}

%

\end{document}